\documentclass[a4paper,12pt]{article}
\usepackage{epsfig}
\newcommand{\beq}{\begin{equation}}

\newcommand{\eeq}{\end{equation}}

\newcommand{\beqa}{\begin{eqnarray}}
\newcommand{\eeqa}{\end{eqnarray}}
\newcommand{\ba}{\begin{array}}
\newcommand{\ea}{\end{array}}

\newcommand{\ben}{Bender {\em et al} (1977)}

\def\fnote#1{\footnote}

\def\square{\hbox{\vrule\vbox{\hrule\phantom{o}\hrule}\vrule}}

\newtheorem{teo}{Theorem}

\newcommand{\be}{\begin{equation}}

\newcommand{\ee}{\end{equation}}

\newcommand{\bt}{\begin{teo}}

\newcommand{\et}{\end{teo}}

\newcommand{\s}{\sigma}
\newcommand{\ii}{{\rm i}}

\begin{document}

\begin{center}
\epsfig{file=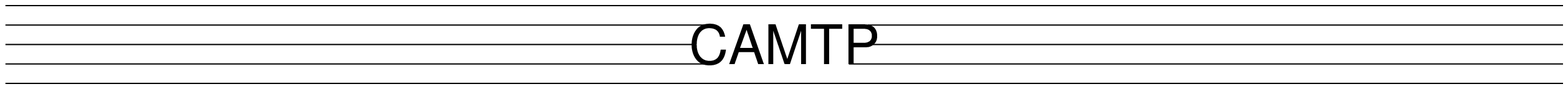,height=8mm,width=\textwidth}\\[2mm]
\end{center}

\begin{flushright}

Preprint CAMTP/99-3\\

September 1999\\

\end{flushright}

\vskip 0.5 truecm
\begin{center}
\Large

{\bf   Some properties of  WKB series }\\

\vspace{0.25in}

\normalsize

Marko Robnik\footnote{e--mail: robnik@uni-mb.si}
and
Valery G. Romanovski\footnote{e--mail: valery.romanovsky@uni-mb.si}

\vspace{0.3in}

Center for Applied Mathematics and Theoretical Physics,\\

University of Maribor, Krekova 2, SI-2000 Maribor, Slovenia\\

\end{center}
\vspace{0.3in}

\normalsize
{\bf Abstract.} We investigate some properties of the  WKB
series for arbitrary analytic potentials and  then specifically
for potentials $x^N$ ($N$ even), where more explicit formulae
for the WKB terms are derived. Our main new results are: (i) We
find the explicit functional form for the general WKB terms $\sigma_k'$,
where one has only to solve a general recursion relation for the
rational  coefficients. (ii) We give a systematic algorithm
for a dramatic simplification of the integrated WKB terms
$\oint \sigma_k'dx$ that enter the energy eigenvalue equation.
(iii) We derive almost explicit formulae for the WKB terms for the
energy eigenvalues of the homogeneous power law potentials
$V(x) = x^N$, where $N$ is even.
In particular, we    obtain  effective algorithms to compute and reduce
the  terms of these   series.


PACS numbers: 03.65.-w, 03.65.Ge, 03.65.Sq

Submitted to {\bf Journal of Physics A: Mathematical and General}

\newpage

\section{Introduction}

Semiclassical approximations of the solutions to the wide variety
of wave equations in mathematical physics are well known since
at least the nineteenth century, mainly based on the ideas
of short wavelength approximations. Their applications became
of extreme importance in solving the energy eigenvalue problem
of the stationary Schr\"odinger equation. In fact, this has been
done by the famous EBK quantization, initiated by Bohr and Sommerfeld
for the separable potentials, generalized by Einstein to the
integrable potentials (by quantizing the classical invariant
tori), and completed by Maslov (by taking into account  the
phase corrections at the classical turning points and caustics
in more than one degree of freedom). This is known under the
name torus quantization. Einstein's work was in fact completed
in 1917 even before the quantum mechanics was discovered, and
it has been later realized that it is (after Maslov
phase corrections are taken into account) precisely the
leading order semiclassical approximation to the exact solution.
Einstein knew from the communications with Henri Poincar\'e
that the integrability in classical Hamiltonian systems
is exceptional, and thus that nonintegrability and nonexistence
of invariant tori is the generic rule. How to proceed in
such case was a mystery for him. It is later through the works
of Gutzwiller around 1970 (see his book Gutzwiller 1990) that we
got an answer to this question, which is a systematic
approach to represent the leading semiclassical approximation
as a sum over contributions associated with classical periodic
orbits, the so called Gutzwiller theory. All these findings
were of crucial heuristic, qualitative and to  large extent
also quantitative value and importance especially in the course
of studying the quantum chaos, which is the study of the quantum
signatures of classical chaos.

On the other hand
in 1-dim systems more can be done and indeed there is a  quite well 
developed theory (see   e.g. Delabaere {\it et al} (1997), Balian {\it et al} (1979), Voros  (1983) and references therein).
Nevertheless if we ask for explicit results 
not so  much is known.  However there are  the systematic
semiclassical procedures in the one dimensional potentials,
which are of course always integrable, and where the leading
torus approximation is only precisely the first term in a certain
WKB series expansion, where each term can be obtained recursively,
in principle, and where in certain cases the series can be
summed and it can be shown to yield precisely the exact result.
In this regard a very important classic and historic  paper
is by Bender {\em et al} (1977). See also (Robnik and Salasnich
1997a,b) and Romanovski and Robnik (1999) and (Salasnich and Sattin
1997). The Schr\"odinger eigenvalue problem is very difficult
even for general one dimensional potentials and any analytically
tractable WKB approximations and exact results are of extreme
importance in qualitative and quantitative understanding
of general solutions. Also, they provide a basis for the
many dimensional integrable cases, where almost  nothing is known
beyond the leading Einstein-Maslov term. In the following
we shall use the terminology  "WKB series expansion" as the
synonym of the "semiclassical expansions" which we clearly define
below.

Although at present the WKB theory is very deeply  developed and its methods
 are very 
important for many applications, there are only  a few works where  the problem 
of effective calculation of the terms of WKB expansions is discussed. In this
direction a pioneering  paper is by \ben,  
where the authors investigated the structure of the terms of WKB expansions 
and also applied the methods to compute the eigenvalues of the potential $V(x)=x^N$ ($N$ even).  In our present  paper we perform  further study of  the problem of effective computation of WKB series started in \ben. We obtain new  recurrence formulae for WKB terms 
for arbitrary analytic potentials and for the polynomial potential $V(x)=x^N$. The known  algorithms are very laborious, because they involve operations of  differentiation and  collection of similar terms in polynomials, which  are extremely  time-consuming as the order increases,  even when modern computer algebra systems are used.
In contradiction, in computing by means of  our recurrence formulae one does only arithmetic operations with rational numbers, as we have explicit formulae for the WKB terms, except for the numerical coefficients.  

Using the obtained  functional form for the general WKB terms $\s'_k$ we  give a systematic algorithm
for a dramatic simplification of the integrated WKB terms
$\oint \sigma_k'dx$ that enter the energy eigenvalue equation,
by showing systematically that the idea of Bender {\em et al} (1977)
can be fully implemented resulting in the maximum possible
simplification of the integrated terms. This is achieved by
finding terms of the integrand which are complete derivatives
of some function and thus give zero contribution when integrated
round a closed integration contour in the complexified
coordinate plane $x$.
 We also  derive almost explicit formulae for the WKB terms for the
energy eigenvalues of the homogeneous power law potentials
$V(x) = x^N$, where $N$ is even.

These results go substantially beyond the results of Bender {\em et al}
(1977) and indeed it should be emphasized that the main algebraic
ideas behind our present work are due to certain remarkable
similarities between our present problems and those involved in
calculating the normal forms and Lyapunov focus
quantities in the power law differential
equations (of one degree of freedom) and  maps (see Romanovski  1993,
Romanovski and Rauh 1998). There are also some  common
features between the problem of reduction of the coefficients $\s'_k$
of the WKB series and the problem of finding a basis of the ideal
of Lyapunov focus quantities (the so-called local 16th Hilbert
problem, see Romanovski 1996). Some introduction to the WKB method can be found in (Bender and Orszag 1978).

We consider the two-turning point eigenvalue problem for the
one-dimensional
 Schr\"o\-dinger equation
\be   \label{sch}
 [- {\hbar ^2} \frac {{\rm d}^2}{{\rm  d} x^2}+V(x)]\psi (x)=E\psi(x).
\ee

We can always write the wavefunction as
\be
\psi(x) =\exp \left\{\frac 1\hbar \sigma (x)\right\}
\ee
where the phase $\sigma (x)$ is a complex function that satisfies the
differential equation
\be \label{w1}
\sigma '^2(x)+ \hbar  \sigma '' (x)=
(V(x)-E)\stackrel{def}{=} Q(x).
\ee
The WKB expansion for the phase is
\be \label{w2}
\sigma(x)=\sum_{k=0}^{\infty}
  \hbar^k \sigma_k (x).
\ee
Substituting (\ref{w2}) into (\ref{w1}) and comparing
like powers of  $\hbar$ gives the recursion relation
\be \label{w3}
\s _0'^2 = Q(x),\ \ \ \ \ \s'_{n}=-\frac 1{2\s'_0}
(  \sum_{k=1}^{n-1} \s' _k \s ' _{n-k}+\s_{n-1}'').
\ee

Computing few first functions $\s'_k$ by means of the recurrent formula
we get
\be
\s'_0=-\sqrt{Q(x)}, \ \ \ \s'_1={\frac{-Q'(x)}{4\,Q(x)}},
\ee
\be
\s'_2=
{\frac{5\,{{Q'(x)}^2} - 4\,Q(x)\,Q''(x)}{32\,{{Q(x)}^{{\frac{5}{2}}}}}},
\ee
\be
 \s'_3=
{\frac{-15\,{{Q'(x)}^3} + 18\,Q(x)\,Q'(x)\,Q''(x) -
4\,{{Q(x)}^2}\,Q^{(3)}(x)}
   {64\,{{Q(x)}^4}}}
\ee
and
\begin{eqnarray}
\s'_4=
(1105\,{{Q'(x)}^4} - 1768\,Q(x)\,{{Q'(x)}^2}\,Q''(x) +
     448\,{{Q(x)}^2}\,Q'(x)\,Q^{(3)}(x) + \\ 304
{{Q(x)}^2}\,{{Q''(x)}^2} -
64\,{Q(x)}^3\,Q^{(4)}(x)  })/
     {2048\,{{Q(x)}^{{\frac{11}{2}}}}.\nonumber
\end{eqnarray}

For the analytical  potential $V(x)$ 
the following quantization condition is known 
(see Dunham (1932), Fr\"oman and Fr\"oman (1977),  Fedoryuk (1983)):
\be \label{gin}
\frac 1{2\ii}\oint_\gamma \sum_{k=0}^{\infty}
\hbar^k  \sigma'_k(x)  dx  = \pi n_q \hbar,
\ee
where $n_q \ge 0$ is  an integer number
and $\gamma $ is a contour surrounding the turning points on the real
axis.
This relation  is an equation with respect to $E$ and using it one can
find the asymptotic of the eigenvalues $E_n(\hbar)$ (see
 e.g.  Balian {\it et al} (1979),  Fedoryuk (1983)  and references therein). In some
cases the series (\ref{gin}) can be summed exactly
(see  Bender {\em et al} (1977), Robnik and
Salasnich (1997a,b), Romanovski and Robnik (1999), Salasnich and
Sattin (1997)).

The zero-order term of the WKB expansion is given by
\be
\frac 1{2 \ii} \oint _\gamma { d} \sigma_0 = \int { d} x
\sqrt{E-V(x)},
\ee
the first odd term is
\be
 \frac \hbar{2 \ii}
  \oint _\gamma{ d } \sigma_1 =-\frac{\pi \hbar}2
\ee
and to find the higher order terms we need to compute
 the functions $\s'_k$ using the recursion relation (\ref{w3}).
We note that the odd-order terms (except the  first order for
$\sigma'_1$) yield integrals that vanish exactly, because,
as it follows from the results of Fr\"oman (1966), the 
functions $\s'_{2k+1}$ are total derivatives.

\section{An algorithm for computing  $\s'_k$}
We will look for  a general  formula for the functions $\s'_k$.
As  it is known one of the most powerful tools for investigation of
recurrence relations is the method of  generating functions (see e.g.
Graham {\it et al} 1994). The most widely used in combinatorics
generating functions are the  ones with  a single variable,
for example, a generating function for the sequence $\{g_0, g_1,
g_2,\dots \}$
is the formal series
\be
G(z)=\sum_{n\ge 0} g_n z^n.
\ee
For a multi-index sequence $\{g_{(y_1,\dots,y_m))}\}_{ y_i \in
{\bf N}}$, where ${\bf N}$ is the set of non-negative integers, a generating function  has  the form
\be
G{(z_1,\dots,z_m)}=\sum_{y_i\ge 0} g_{(y_1,\dots,y_m)} z_1^{y_1}\dots
z_m^{y_m}.
\ee
Moreover, we can also consider a sequence, where every term has only
finite number of indices, but the total number of indices is unbounded
(e.g. $ g_{(y_1)}, g_{(y_1,y_2)},$ $  g_{(y_1,y_3,\dots,y_n)},\dots ) $
with the generating function
\be \label{gf}
G{(z_1,z_2,\dots)}=\sum_{\gamma \in {M}} g_{\gamma} \bar z^{\gamma},
\ee
where $M=\cup_{k=1}^\infty {\bf N}^k$, $\bar z=(z_1,\dots,z_s) $ and
$\bar z^{\gamma}=z_1^{\gamma_1}\dots z_s^{\gamma_s}$. Thus in the last
case $ G(z_1,z_2,\dots)$ is an element of the
ring of formal power series in the infinite number of variables, $
z_1,z_2, \dots .$

We now apply the method of generating functions to computing of the WKB
expansion for the phase.
With any vector $\nu=(\nu_1,\nu_2,\dots, \nu_l)$ we associate the
operator
\be \label{lk}
L(\nu)= 1\cdot \nu_1+2\cdot \nu_2 +\dots + l \cdot \nu_l,
\ee
where the  vector $\nu$ runs through whole $M$.

We will show that the functions $\s'_m$  are of the form
\be \label{sk}
\s'_m=\sum_{\nu: L(\nu)=m} \frac{U_\nu
Q^{m-|\nu|}Q^{(\nu)}}{Q^{\frac{3m-1}{2}}},
\ee
where for a vector $\nu=(\nu_1,\dots, \nu_l) $ we denote   $Q^{(\nu)}=
(Q')^{\nu_1}(Q'')^{\nu_2}\dots (Q^{(l)})^{\nu_l},$
$|\nu|=\nu_1+\dots+\nu_l$ and the coefficients $U_\nu $ satisfy the
recurrence
relation
\be \label{u}
U_\nu=\frac 12 \sum_{\mu,\theta\ne 0, \mu+\theta=\nu} U_\mu U_\theta +
\frac{(4-L(\nu)-2|\nu|)U_{(\nu_1-1,\nu_2,\dots,\nu_{l})}}4+\sum_{i=1}^{l-1}
\frac{(\nu_i+1)U_{\nu(i)}}2,
\ee
where $U_0=-1$
and we put
$U_\gamma=0$ if among the coordinates of the vector $\gamma$ there is
negative one, and we denote  by  $\nu(i)$ $ (i=1,\dots,l-1)$ the vector
$(\nu_1, \dots, \nu_{i}+1,\nu_{i+1}-1,\dots,\nu_{l}).$

It should be mentioned that the complexity of functions $\s'_n$ increases rapidly with the order $n$, and it is remarkable that applying our almost explicit 
 formulae
(\ref{sk}), (\ref{u}) we can go much further  than by using the well-known recursion relation (\ref{w3}) (see the  Appendix). 

Note that the number of solutions $\nu$ of the equation
\be \label{19}
L(\nu)=m,
\ee
where $m>0$ and $\nu$ is a vector with non-negative coordinates, equals
the number of partitions $p(m)$ of the integer $m$ (there exists a
theory and a formula for $p(m)$; see e.g. Andrews (1976)).
Therefore as a corollary of formula (\ref{sk}) we find  that the number
of terms
of the function $\s'_k$ cannot exceed $p(m)$. This fact was for the
first time observed by Bender {\it et al} (1977).

We prove the  formulae (\ref{sk}), (\ref{u}) by induction on
$m$. Indeed, for $m=1,2$ the statement holds.
Let us suppose that it is true  for all $m$ less than $k$. Then for
$m=k$ we get
\be \label{10}
\frac{\s'_i\s'_{k-i}}{2 \s'_0}=
\sum_{\theta,\mu: L(\theta)=i, L(\mu)=k-i} - \frac{U_\theta U_\mu
Q^{k-|\mu|-|\theta|} Q^{(\mu+\theta)}}{2 Q^{\frac{3k-1}{2}}}.
\ee
Obviously,
\be \label{t1}
L(\theta)=i, L(\mu)=k-i \Rightarrow L(\theta+\mu)=k,
\ee
and
here we assume that the dimensions of vectors $\theta$ and $\mu$ are the
same,
if not we simply extend the dimension of the smaller one by putting
zeros for the excess coordinates.
On  the other hand, taking into account that all coefficients of the
operator
(\ref{lk}) are positive, it is easy to verify that
\be \label{t2}
L(\theta + \mu) =k \  \Rightarrow L(\theta)=j, L(\mu)=k-j, \ 0\le j\le k
\ee
(this property is the crucial one for the presented method).
Thus (\ref{10})--(\ref{t2}) yield
\be \label{11}
\sum_{i=1}^{k-1} \frac{\s'_i\s'_{k-i}}{2 \s'_0}=
\sum_{\theta,\mu: L(\theta+\mu)=k} - \frac{U_\theta U_\mu
Q^{k-|\mu|-|\theta|} Q^{(\mu+\theta)}}{2 Q^{\frac{3k-1}{2}}},
\ee
i.e. we get an expression of the form (\ref{sk}).

For the last term of the recurrence formula (\ref{w3}) we get
\be
\frac{\s''_{k-1}} {2\s'_0} =\sum_{\mu: L(\mu)=k-1} -  U_\mu
\left[ \frac{ (2-k-2 |\mu|) Q^{k-|\mu|-1}Q^{(\mu_1+1,\mu_2,\dots,
\mu_{k-1})}}
{4 Q^{\frac{3k-1}2} }+ \frac{Q^{k-|\mu|-1}(Q^{(\mu)})'}{2
Q^{\frac{3k-3}2}} \right]
\ee
Note that for the vector $\mu=(\mu_1,\dots,\mu_{k-1},0)$
\be  \label{13}
[Q^{(\mu)}]'=\sum_{i=1}^{k-1} \mu_i Q^{(\hat \mu(i))},
\ee
where we denote by $\hat \mu(i)$ \  $(i=1,\dots,k-1)$ the vector
$(\mu_1, \dots, \mu_{i}-1,\mu_{i+1}+1,\dots,\mu_{k})\ (i=1,\dots,k-1).$
It is readily seen that $L(\hat \mu(i))=L(\mu)+1=k,$ therefore,
formulae (\ref{w3}), (\ref{11})--(\ref{13}) yield that (\ref{sk}),
(\ref{u})
 hold. \square

Using the recurrence relation (\ref{u}) we can obtain  the  differential
equation
for the generating function of the sequence $U_\nu$
\be
U(\bar z)=U(z_1,\dots)=\sum_{\nu\in M} U_{\nu} \bar z^{\nu}.
\ee
Let us rewrite (\ref{u}) in the form
\begin{eqnarray} \label{unn}
U_{(\nu_1,\dots,\nu_l)}=\frac 12 \sum_{\mu,\theta\ne 0, \mu+\theta=\nu}
U_{\mu} U_{\theta} +
 U_{(\nu_1-1,\nu_2,\dots,\nu_l)}-\frac 34 \nu_1
U_{(\nu_1-1,\nu_2,\dots,\nu_l)}
-\\ \frac 14 \sum _{i=2}^l (i+2) \nu_iU_{(\nu_1-1,\nu_2,\dots,\nu_l)}
+\sum_{i=1}^{l-1}\frac{(\nu_i+1)U_{\nu(i)}}{2}-[\nu=0],\nonumber
\end{eqnarray}
where $[\alpha =\beta]$ denotes the function, which equals 1 if
$\alpha=\beta $ and 0 otherwise.

Using obvious properties of generating functions (see e.g. Graham {\it
et al} 1994) we get from (\ref{unn})
\be \label{upart}
U=\frac 12 (U+1)^2+z_1 U-\frac 34 z_1 (z_1U)'_{z_1}-\sum_{i=2}^l \frac
{i+2}4 z_1 z_i U'_{z_i}+
\frac 12 \sum_{i=1}^{l-1}z_{i+1} U'_{z_i}-1.
\ee
It means if we fix any integer $l$ and, therefore, the variables
$z_1,\dots,z_l$, then the function
\be
\hat U(z_1,\dots,z_l)= U(z_1,\dots,z_l,0,0,\dots)
\ee
 is the solution of the equation (\ref{upart}) with
the initial conditions
\be \label{uh}
\hat U(0)=-1, \hat U'_{z_i}(0)=-\frac 1{2^{i+1}}
\ee
(we get the initial conditions  from (\ref{u}) taking into account that
 $U_{(0,\dots,0,1)}=-\frac 1{2^{i+1}}$ for the vectors with the only
$i$th coordinate different from zero). So,    the coefficients $U_\nu$
that  enter
 the functions $\s'_k$ (\ref{sk})
are precisely  the coefficients  of the Taylor expansion of the
function $\hat U$ defined by the  differential equation  (\ref{upart})
with the initial conditions  (\ref{uh}).

Coefficients of the form $U_{(n,0,\dots,0)}$ depend on the coefficients
of the same form.
Therefore we get from (\ref{upart}) that
 the function $U(z)=\sum_{n=0}^\infty U_n z^n   $ satisfies the
differential equation
\be
U=\frac 12 (U+1)^2+ z U-\frac 34 z (zU)'_z-1,
\ee
which is the Riccatti equation
\be \label{ric}
3 z^2 U'_z-2 U^2 -z U+2=0.
\ee

Note that as an immediate  corollary of formula  (\ref{sk}) we get that
for the harmonic oscillator, i.e. when $Q= {x^2}-E$, the WKB series
(\ref{gin}) terminates after the first two terms, namely,
\be \label{ho}
\oint_\gamma {\rm d}\sigma_k=0
\ee
for all $k\ge 2.$ Indeed, in this case (\ref{sk}) yields
\be
\s'_{m}=\sum_{i=0}^{[m/2]} \frac{U_{(m-2i,i)} 2^{m-i} x^{m-2i}}
{\sqrt{x^2-E}^{3 m-1-2i}},
\ee
where $[m/2]$ stands for the integer part of $m/2$.

It is obvious
\be
{\rm Res}_\infty \frac{ x^{m-2i}}{\sqrt{x^2-E}^{3m-1-2i}}=0
\ee
for all $m>1$. Therefore (\ref{ho}) takes place.

\section{An algorithm for the simplification of the functions $d \s _k$}

It was pointed out by Bender {\it et al} (1977) that  the functions
d$\s_k$
can be dramatically simplified by adding and subtracting total
derivatives
(obviously, such operation does not change the   integrals (\ref{gin})).
They also carried out numerical experiments to obtain different
simplifications of these  functions. It is easily seen that the formula
(\ref{sk})
provides an effective  way to reduce  the number of terms in the
expressions for the  functions $d \s _k.$

We will look for a function of the form
\be \label{pk}
P_k=\sum_{\mu: L(\mu)=k} \frac{W_\mu
Q^{k-|\mu|}Q^{(\mu)}}{Q^{\frac{3k}{2}}},
\ee
where $W_\mu$ are to be determined,  such that
\be \label{eqp}
 \frac{d}{dx} P_{k} =\s'_{k+1}
\ee
By  comparing coefficients of
$
 \frac{ Q^{k+1-|\nu|}Q^{(\nu)}}{Q^{\frac{3k+2}{2}}},
$
in  both parts of
(\ref{eqp}) we get the system
\be \label{sys}
\frac{3-L(\nu)-2 |\nu|}2 W_{(\nu_1-1,\nu_2,\dots,\nu_{l})}
+ \sum_{i=1}^{l-1}(\nu_i+1) W_{\nu(i)} = U_{\nu},
\ee
where $\nu=(\nu_1,\dots,\nu_l)$ runs through the whole set of  solutions
of equation (\ref{19})
and $U_\nu$ are defined by the recurrence relation (\ref{sk}),(\ref{u}).

Thus to simplify the function $\s'_{k+1}$ one can solve the system
(\ref{sys})
 of $p(k+1)$  equations in $p(k)$ variables $W_\mu$.
For example, to simplify $\s'_4$ we write down the corresponding system
(\ref{sys}) and get
\be  \label{sys4}
 \begin{array}{lllll}
U_4 & = &- \frac 92 W_3& &\\ U_{(2,1)} & = & 3 W_3 -& \frac 72
W_{(1,1)}&\nonumber \\  U_{(1,0,1)}& =& &W_{(1,1)}-&\frac 52
W_{(0,0,1)}\\
U_{(0,2)} & =& &  W_{(1,1)}  &
\\  U_{(0,0,0,1)} & = & & &W_{(0,0,1)}
\end{array}
\ee
where $U_4=1105/2048, U_{(2,1)}= -1768/2048, U_{(0,2)}=304/2048,
 U_{(1,0,1)}=448/2048,$ $ U_{(0,0,0,1)}= -64/2048.$
We see that  the matrix, corresponding to the first three equations
is the triangular  one, so   we can kill three terms in the expression
for $ {\rm d}\s_4.$
 Computing we get
\be
W_3= -\frac{1105}{9216}, W_{(1,1)}= \frac{ 221}{1536},
      W_{(0,0,1)}= -\frac{23}{768}.
\ee
Hence
\be
\s'_4-\frac d{dx} P_3=\frac{
7 Q''(x)^2 - 2 Q(x) Q^{(4)}(x)}
  {1536 Q(x)^{7/2}},
\ee
in accordance with Bender {\it et al} (1977).

Let us denote by $\tilde p(k)$ the number of partitions of $k$ which
contain at least one 1. Obviously, $p(k)=\tilde p(k+1)$. It appears that
the optimal strategy
to simplify $\s'_{k}$ is as follows. In system (\ref{sys})  where
$L(\nu)=k$ we consider the equations with $\nu$ such that $\nu_1\ne 0.$
There are $\tilde p(k)=p(k-1)$ such equations and according to
(\ref{pk})
we have exactly $p(k-1)$ variables. It turns out that
we can always  write the systems with $U_\nu$ such that
$\nu_1\ne 0$ in the triangular form (like system (\ref{sys4})).
To see this we set the following order on vectors of ${\bf N}^l$:
we say that $\nu< \mu$ if the first nonzero entry from the left in
$\mu-\nu$ is positive (this order is known in computational algebra as
the lexicographic one). Then  if we write down the  equations of  the
system (\ref{sys}), corresponding to $U_{\nu^{(1)}}, U_{\nu^{(2)}},
\dots $ in the decreasing order
$\nu^{(1)}>\nu^{(2)},\dots   $
and the variables $W_\mu$ in these
equations also in the decreasing order then  we get  that the matrix
corresponding to
the first $p(k-1)$ equations is the triangular $p(k-1)\times p(k-1)$
matrix
(because $\nu(i)>(\nu_1-1,\nu_2,\dots,\nu_l)$ for  $1\le i\le l-1$).
Moreover, the diagonal elements of the matrix are equal to
$({3-L(\nu)-2|\nu|})/2$ and, therefore,  are different from zero.
Hence, we get that after the simplification $\s'_{2n} $
contains at most $p(2n)-p(2n-1)$ terms.

To end this section we show that in some cases we can replace
calculation of
contour integral by computing  a Riemann integral, namely we shall show
that
 formula (\ref{43}) below applies.

First we note that taking into account that $Q= V(x)-E$ we can write for
even $ m $ formula (\ref{sk}) in the form
\be \label{ske}
\s'_m=\sum_{\nu: L(\nu)=m} \frac{2^{\frac m2-1+ |\nu|}\ \ii}{ (m-3+2
|\nu|)!!} \frac {\partial ^{\frac m2-1+|\nu|}}{\partial E ^{\frac
m2-1+|\nu|} } \frac{U_\nu V^{(\nu)}}{\sqrt{E-V}}.
\ee

Let us now  suppose that $Q= V(x)-E$, where $V(x)$ is an analytic  function
with
one minimum and $V'(x)\ne 0$, if $x\ne 0$.  We will show that
\be \label{43}
\oint {\rm d}\s_m=2 \sum_{\nu: L(\nu)=m} \frac{2^{\frac m2-1+
|\nu|}\ \ii}{ (m-3+2|\nu|)!!} \frac {\partial ^{\frac m2-1+|\nu|}}{\partial E
^{\frac m2-1+|\nu|} } \int_{x_1}^{x_2} \frac{U_\nu
V^{(\nu)}}{\sqrt{E-V}} dx,
\ee
where  $V(x_1)=V(x_2)=E_0, \ x_1< x_2. $ Taking into account (\ref{ske})
it is easy to see
that to  prove (\ref{43}) it is sufficient to show that
\be \label{dp}
\oint_\gamma \frac {\partial ^{\frac m2-1+|\nu|}}{\partial
E ^{\frac m2-1+|\nu|} } \frac{ V^{(\nu)}}{\sqrt{E-V}} dx=
2  \frac {\partial ^{\frac m2-1+|\nu|}}{\partial
E ^{\frac m2-1+|\nu|} }\int_{x_1}^{x_2} \frac{ V^{(\nu)}}{\sqrt{E-V}}
dx.
\ee
Note that  due to the theorem on differentiation upon a parameter (see
e.g. Sidorov {\it et al} 1976) if
\be
F(E)=\oint_\gamma f(x,E) dx
\ee
and\\
1) $\gamma$ is a finite piecewise-smooth curve;\\
2) the function $f(x,E)$ is continuous with respect to  $(x,E) $
for $ x\in \gamma, E\in D$, where $D$ is a domain of the complex
plane;\\
3) for every fixed $x\in \gamma$ the function
$f(x,E)$ is analytic upon  $E$ in $D$,
then $F(E)$ is analytic in $D$ and
\be
F'(E)=\oint_\gamma \frac{\partial f(x,E)}{\partial E} dx,
\ee
for $E\in D$.

Let us cut the complex plane between the turning  points $x_1$ and $x_2
$
to get a single-valued function and fix the contour
$(x_1+\rho,x_2-\rho)\cup
c_1\cup (x_2-\rho,x_1+\rho )\cup
c_2, $ where   $\rho$ is small $c_1, c_2$ are circles : $c_1=x_2+\rho
e^{\ii t} $, $c_2
=x_1+\rho   e^{\ii t} $ and  $t\in [0,2\pi]$.  Then the conditions 1)-3)
are satisfied (with $D$ being a small neighborhood of $E_0$).
Therefore
\begin{eqnarray}
\oint_\gamma \frac {\partial ^{\frac m2-1+|\nu|}}{\partial
E ^{\frac m2-1+|\nu|} } \frac{ V^{(\nu)}}{\sqrt{E-V}} dx=
\frac {\partial ^{\frac m2-1+|\nu|}}{\partial
E ^{\frac m2-1+|\nu|} }\oint_\gamma  \frac{ V^{(\nu)}}{\sqrt{E-V}} dx=\\
  \frac {\partial ^{\frac m2-1+|\nu|}}{\partial
E ^{\frac m2-1+|\nu|} } ( 2 \int_{-x_0+\rho}^{x_0-\rho} \frac{
V^{(\nu)}}{\sqrt{E-V}} dx
+\oint _{c_1} \frac{ V^{(\nu)}}{\sqrt{E-V}} dx+ \oint _{c_2} \frac{
V^{(\nu)}}{\sqrt{E-V}} dx).\nonumber
\end{eqnarray}
Let us denote
\be
g(E)=\oint_{c_1}  \frac{ V^{(\nu)}}{\sqrt{E-V}} dx.
\ee
 Due to the theorem mentioned above $g(E)$ is analytic in a neighborhood
of
$E_0$, therefore
\be
g(E)\approx g(E_0)+g'(E_0) (E-E_0).
\ee
Noting that
\be
|g(E_0)|=
\left|\oint _{c_1} \frac{ V^{(\nu)}}{\sqrt{E_0-V}} dx  \right|
=\left|\oint_{c_1} \frac{ V^{(\nu)}}{\sqrt{V'(x_1)(x-x_1)+\dots }}
dx\right| <const \  \rho^{1/2}
\ee
we obtain that $g(E)\to 0$ when  $\rho \to 0, E\to E_0$.
It means that formula (\ref{dp}) indeed  holds. This formula has been
found
useful for  computing  the WKB series for Coulomb potential
and the potential $V(x)=U_0/{\rm cos}^2(\alpha x)$ (Robnik and Salasnich
1997a,b).

\section{Potentials of the form $V(x)=x^N.$}

In recent decades  many studies have been devoted to the  investigation of the semiclassical expansions for  
the  potentials of the form $V(x)=x^N$ and important results have been achieved (see e.g. Balian {\it et al} (1979), Voros (1983)  and references therein).  

One of the basic formulae for these potentials was obtained by \ben \  and is as   follows:
\be \label{vb}
\pi(n_q+\frac 12) =E^{1/N+1/2}\sum_{n=0}^{\infty}E^{-n(1+2/N)}a_n(N),
\ee 
where $N$ is an even integer number,
\be
a_n(N)=\frac{{\left( -1 \right) }^
     {n}\,
    2^{1 - n}\,
    {\sqrt{\pi }}\,
    \Gamma(1 + 
      \frac{1 - 
         2\,n}{N}
      )\,P_n(N)}{\left(
       2 + 
       2\,n
       \right) !\,
    \Gamma(
        \frac{3 - 
         2\,n}{2}
       + \frac{1 - 
         2\,n}{N}
      )}
\ee
 and $P_n(N) $ are polynomials of the variable $N$ with integer coefficients. 
The first eight  polynomials $P_n(N)$ where computed using MACSYMA computer algebra program by \ben  \  for the general potential $V(x)=x^N$ and, as it was mentioned, the computation of the eighth polynomial has already faced difficulties. When   $N$ is a fixed integer number then instead of the polynomials $a_n$ we have integers.
The special case of quartic oscillator ($N=4$) has been deeply investigated by Balian, Parisi, Voros and others. 
In this case the expression of the type  (\ref{vb}) is written   (Balian {\it et al} 1979)
in the form 
\be \label{e53}
2\pi(n_q+\frac 12)\hbar =\sum_{n=0}^\infty b_n
\sigma^{1-2n}\hbar^{2n},
\ee 
where
\be
\sigma=E^{3/4} B(3/2,  1/4)
\ee
is the classical action around the closed orbit of energy $E$
(here  $B(x,y)$ is the beta function)  and $b_n$ in this case are rational numbers.
As it is reported by  Balian {\it et al} (1979) with REDUCE language  they were able to compute the first seventeen coefficients $b_n$. Then they had to switch to the ordinary numerical computation and computed $b_n$ up to $n=53$
 (in (Voros 1983) the results of computation up to $n=60$ are presented).  
 
It was mentioned in Balian {\it et al} (1979)  and in \ben \ that the authors do not know any closed form or a simple law for the coefficients $a_n(N)$  and 
$b_n$.
In this section we partially answer this question. Although we also were not able to find any closed form for the functions  $a_n$ (or numbers $b_n$) we obtain a simple recurrence
formula, where   only operations of summation  and multiplication of rational numbers are involved, whereas using the usual way of the above mentioned  papers one needs first to compute  rational functions of the form $f(x,\sqrt{E-x^N})$ and then evaluate contour integrals.
We carried out computer experiments and found out that using our algorithm 
 with Mathematica 4.0 on a  PC with 128 MB RAM  we were able to compute in closed arithmetic form 
the coefficients $b_n$ at least up to $n=190$  for the quartic potential
(see also the Appendix).

It can be  proven (Robnik and Romanovski 2000)
by
induction  using the recursion relation (\ref{w3})  that for the potential $ V(x)=x^N$ 
 the coefficients $\s'_k$\ $(k \ge 1)$ of the WKB expansion
have the form
\be \label{sxn}
\s'_k=-\frac {(-\ii)^{3k-1} x^{-k+N}}{(E-x^N)^{\frac{3k-1}2}}
\sum_{j=0}^{k-1} A_{k-j-1,j} E^{k-j-1} x^{jN},
\ee
where  we choose $
\sqrt{E-x^n}=\ii \sqrt{x^n-E}$, and the coefficients $A_{k-j-1,j}$ of
the monomials $ E^{k-j-1} x^{jN}$ are computed according to the
recurrence
formula
\begin{eqnarray} \label{fa}
\lefteqn{A_{s,l}=\frac 12 \sum_{i=0}^s \sum_{j=0}^{l-1}
 A_{i,j} A_{s-i, l-1-j}+}\\
& &
  \frac{l(2+N)+(2+3N) s-N}4 A_{s,l-1}+\frac {(N-1) l +N-s} 2
A_{s-1,l}\nonumber.
\end{eqnarray}
with  $A_{0,0}=N/4$ and
\be \label{cond}
A_{\alpha,\beta}=0\ {\rm if}\  \alpha<0\ {\rm or}\  \beta<0.
\ee

Using equation (\ref{fa}) we can get the differential equation 
for the generating function of the coefficients $A_{s,l}$,
and in the special case $A_{s,0}$ one can find the explicit formula
(Robnik and Romanovski 2000) 
\be
   A_{s,0}=\frac{N!}{2^{s+2}(N-s-1)!}.
\ee

For even $k, \ (k \leftrightarrow 2k)$  we can write formula (\ref{sxn})
in the form
\be \label{sxnd}
\s'_{2k}=\sum_{j=0}^{2k-1} \frac{\ii 2^{3k}}{(6k-3)!!}
 A_{2k-j-1,j} E^{2k-j-1}
  x^{-2k+(j+1)N} \frac {\partial ^{3k}
}{\partial{E}^{3k}}{(E-x^N)^{1/2}}.
\ee
As above we can replace the integration on a contour with the
integration
between turning points, then, taking into account that
\be
\int_0^a x^{\alpha-1} (a^\theta-x^\theta)^{\beta-1} dx=\frac {a^{\theta
(\beta-1)+\alpha}}{\theta} B(\frac\alpha\theta,\beta),
\ee
where $\alpha,\theta, Re\, \alpha, Re \, \beta >0,$ and $ B$ is the
beta-function,
and noting that
$$
\frac {\Gamma (\frac 32 +s +\frac{1-2k}N+1)}{\Gamma(\frac 32 +s
+\frac{1-2k}N-3 k )}=
$$
\be
(\frac 32 +s +\frac{1-2k}N)(\frac 32 +s +\frac{1-2k}N-1) \dots (\frac 32
+s +\frac{1-2k}N-3 k)
\ee
 we get from (\ref{sxnd})
$$
\oint_\gamma {\rm d} \s_{2k} =2 \int_{-E^{1/n}}^{E^{1/n}} {\rm d}
\s_{2k}=
$$
\be
 \frac {  2^{3k+1}\ii  \sqrt \pi}{(6k-3)!! N} E^{\frac 12+\frac 1N}
\sum_{s=0}^{2k-1}
 A_{2k-s-1,s} E^{-\frac {2k}N-k}
  \frac {\Gamma(\frac{1-2k}N+s+1)}{\Gamma(\frac 32 +s+1-3k+\frac
{1-2k}N)},
\ee
and using the equality $\Gamma(1+z)=z\Gamma(z)$ we  obtain finally the
coefficients of the  WKB expansion
\begin{eqnarray}\nonumber \label{enc}
\oint_\gamma  \s'_{2k} dx=\frac { \ii 2^{3k+1}\sqrt \pi}{(6k-3)!! N}
E^{\frac 12+\frac 1N-\frac{2k}N-k}
\frac {\Gamma(\frac{1-2k}N+1)}{\Gamma(\frac {3-2k}2
+\frac {1-2k}N)} (A_{2k-1,0}\prod_{s=1}^{2k-1} (\frac{3-2k}2+\\
\frac{1-2k}N-s)+\sum_{i=1}^{2k-1}
 A_{2k-i-1,i} \prod_{s=1}^i (s+\frac{1-2k}N)
\prod_{s=1}^{2k-i-1}(\frac{3-2k}2+\frac{1-2k}N-s )),
\end{eqnarray}
where $k\ge 1$ and $ A_{2k-i-1,i}$ are computed according to (\ref{fa})
and
\be
\oint_\gamma \s'_0 dx=\frac{2\ii E^{\frac 12+\frac 1N} \sqrt \pi \Gamma
(1+\frac 1N)}{\Gamma (\frac 32+\frac 1N)}.
\ee

\section{Conclusions}

In this paper we have investigated the  WKB approximations as a
series for arbitrary analytic potentials. Following the classic
paper by Bender {\em et al} (1977) we have achieved the following
results: (i) We find the explicit functional form for the general
WKB terms $\sigma_k'$,
where one has only to solve a general recursion relation for the
{\em numerical  rational} coefficients. (ii) We give a systematic algorithm
for a dramatic simplification of the integrated WKB terms
$\oint \sigma_k'dx$ that enter the energy eigenvalue equation.
(iii)  We derive almost explicit formulae for the WKB terms for the
energy eigenvalues of the homogeneous power law potentials
$V(x) = x^N$, where $N$ is even.
In particular, we    obtain  effective algorithms to compute and reduce
the  terms of these   series.  In computing by means
of these formulae we manipulate only with {\em numbers} and do not need to
collect similar terms of a polynomial, which we must do otherwise
when we use just the recursion formula (\ref{w3}).
Application of the
obtained formulae along with the reduction formula (\ref{sys})
considerably simplifies calculations, especially if we
need to compute high order terms.

\section*{Acknowledgments}
This project was  supported by the Ministry of Science and Technology of
the Republic of Slovenia and by the  Rector's Fund of the University of
Maribor.
VR acknowledges the support of the work by the grant of the Ministry of
Science
and Technology of the Republic of Slovenia and  the Abdus Salam   ICTP
(Trieste) Joint Programme and also the support of  the Foundation of
Fundamental Research of the
Republic of Belarus.

\section*{Appendix}

Here we present the results of computer experiments which we carried out with Mathematica 4.0 on our PC with 450 MHz processor and 128 MB RAM to compare 
the efficiency of our algorithms based on equations (\ref{sk}), (\ref{u})
and  (\ref{fa}),
 (\ref{enc})  with  traditional ones. 
\begin{table}[h]
\caption{\protect
(a) the order $n$ of the coefficient $\s'_n$, (b) the number of terms in the coefficient  $\s'_n$ (equals $p(n)$), (c) CPU time  in seconds   for computing $\s'_n$ by means
of formula (\ref{w3}), (d) CPU  time   for computing $\s'_n$ using
formulae (\ref{sk}), (\ref{u}).
}\begin{center}
\begin{tabular}{rrrrrrrr}
\hline
(a) & (b) & (c)   & (d)  &  (a) & (b) & (c)   & (d)   \\ \hline    
10& 42   &    0.47        &    0.54 & 30& 5604 &  18,542   &    1,402 \\
15& 176  &   5.7        &    5.2   & 34 & 12310 &  114,219   &     4,981\\
20& 627  &   83       &    41      &  35& 14883 &  --    &    6,733 \\  
25& 1958 &    1,509          &    256 & 40& 37338 & -- & 29,940 \\ \hline
\end{tabular}
\end{center}
\end{table}

To compute  $\s'_n$ by  Mathematica using  formula  (\ref{w3}) 
one can  just use it in the form presented in the text, but for computing by means of formulae (\ref{sk}), (\ref{u}) a procedure has to be written.

In the case of the potential $V(x)=x^4$ for computing 
 $\oint \s'_{2n} dx$ one can use formulae (\ref{fa}),
 (\ref{enc}) precisely   in the form presented in the paper, but it is necessary preliminary to define $A_{-1,i}=A_{i,-1}=0$.

In (Voros 1983) the table of results of numerical calculations 
of the numbers 
$$
A_n=\frac{2^
     {- \frac{3}{2}
            - n}\,
    \pi \,b_n\,
    \csc (\frac{
        \left( 3 - 6\,n
          \right) \,\pi }{4})}
    {\Gamma( 
     2\,n-1)}
$$
which should not be confused with our $A_{s,l}$ and where
the numbers $b_n$ are defined in equation (\ref{e53}),    with an estimated accuracy of 34 digits 
are presented and it is mentioned there that the accuracy is not guaranteed 
for $n$ close to 60. Indeed, we found perfect correspondence with the 
table of  (Voros 1983) up to $n=52$, however for larger $n$ there is 
 a disagreement with our  calculations, presented in  Table 2 (digits which differ from those  obtained    by Voros (1983) are underlined).
It should be emphasized that here we calculate $A_n$ in  the exact arithmetic form, which includes rational numbers and the gamma function, but show here the numerical results just for the purpose of comparing them with those of  (Voros 1983).

\begin{table}
\begin{center}\caption{\protect $V(x)=x^4$}
\begin{tabular}{cc}
\hline
n  &  $A_n$      \\ \hline  

53 & 0.9949789006826957939983352522204\underline{201}  \\
54 & -0.99507337304487214360383324779308\underline {91}\\
55  &   0.995164356634546100730933541912349\underline{5}    \\ 
56 & -0.9952520411721716125492612513022\underline{471}   \\ 
57 &  0.9953366028692452324220366394\underline{796240}    \\ 
58 &     -0.9954182056094612326949059\underline{324394196}     \\ 
59 &    0.9954970020080866308729\underline{698916050341}     \\ 
60& -0.9955731343639551368199525397\underline{204894} \\ \hline   
\end{tabular}
\end{center}
\end{table}

We computed the coefficients $b_n$ in exact arithmetic form  according to 
formulae  (\ref{fa}),
 (\ref{enc})  up to $n=190$ and it took 143555 sec CPU time to do so.
  It was possible to continue computations according to the  memory capacity,
however the computations became too time consuming. 
As   it is reported  by Balian {\it et al} (1979)  with REDUCE language they 
were  able to go up to $n=16$.


\newpage


\begin{thebibliography}{99}
\bibitem{}
Andrews G E 1976 {\it The Theory of Partitions} (Reading, Massachusetts:
Addison-Wesley)
\bibitem{}
Balian R, Parisi G and Voros A 1979  {\it Feyman Path Integrals"}
Springer Lect. Notes Physics {\bf 106} 337
\bibitem{}
Bender C M, Olaussen K and Wang P S 1977 Phys. Rev. D {\bf 16} 1740
\bibitem{}
Bender C M and Orszag S A 1978 {\it Advanced Mathematical Methods for Scientists and Ingineers } (Auckland: McGraw-Hill)
\bibitem{}
Delabaere E, Dillinger H and Pham F 1997 J. Math. Phys. {\bf 38}(12) 6126  
\bibitem{}
Dunham J L 1932 Phys. Rev. {\bf 41} 313
\bibitem{}
Graham R L, Knuth D E and Patashnik O 1994  {\it Concrete Mathematics}
(New York: Addison-Wesley Publishing)
\bibitem{}
Gutzwiller M C 1990 Chaos in Classical and Quantum Mechanics (Berlin:
Springer)
\bibitem{}
Fedoryuk M V 1983 {\it Asymptotic methods
for Linear Ordinary Differential Equations} (Moscow: Nauka; in Russian)
\bibitem{}
Fr\"oman N 1966  Ark. f\"or Fysik {\bf 32} 541
\bibitem{}
Fr\"oman N and Fr\"oman P O 1977 J. Math. Phys. {\bf 18} 96
\bibitem{} Robnik, M. and Romanovski, V. On the semiclasical  expansion for 1-dim $x^N$ potentials, to appear   in the Proceedings of the 4th School/Conference
"Let's Face Chaos through Nonlinear Dynamics", Maribor, Slovenia, June/July
1999, ed. by M.Robnik et al, Prog. Theor. Phys. Suppl. (Japan). No.139, 2000.
\bibitem{}
Robnik M and Salasnich L 1997a J.Phys. A: Math Gen {\bf 30} 1719
\bibitem{}
 Robnik M and  Salasnich L
 1997b  J.Phys. A: Math Gen {\bf 30} 1711
\bibitem{}  Romanovskii V G 1993
Differentsial'nye  Uravneniya,   {\bf 29}
910 (in Russian); 1993  Differential Equations, {\bf 29},
 782 (English translation)
\bibitem{RM} Romanovski V G 1996
{\em Ideals in  monoid rings and the 16-th Hilbert problem}
Submitted to   J. Symbolic  Computation
\bibitem{}
Romanovski V G and
Rauh A 1998
Dynamic Systems and Applications,   {\bf 7}
529
\bibitem{}
 Romanovski V and Robnik M 1999 {\em On WKB Series for the Radial
Kepler Problem} Preprint CAMTP/99-2, April 1999, to appear in
{\em Nonlinear Phenomena in Complex Systems (Minsk)} {\bf 2} (1999).
\bibitem{}
 Salasnich L and Sattin F 1997 J.Phys. A: Math Gen {\bf 30} 7597
\bibitem{}
Sidorov Yu V, Fedoryuk M V and Schabunin M I 1976 {\it Lectures on
Theory of Funcions of a Complex Variable} (Moscow: Nauka; in Russian)
\bibitem{}
Voros A 1983 Ann. Inst. H.Poincare {\bf 38} 211 
\end{thebibliography}
\end{document}